\documentstyle[12pt,epsf]{article}
\begin{document}

\begin{titlepage}
\renewcommand{\thefootnote}{\fnsymbol{footnote}}

\begin{center}
~hep-ph/9808226 \hfill
TPI-MINN-98/13-T~\\
\hfill  UMN-TH-1713-98~~~\\
\end{center}

\vspace{.3cm}

\begin{center} \Large
{\bf Heavy flavor decays and OPE in two-dimensional 't Hooft
  model\footnote
{Talk presented at the 3rd workshop on
Continious Advances in QCD, University of Minnesota,
Minneapolis, April 16-19, 1998;
to appear in the Proceedings
}
}
\end{center}
\vspace*{.3cm}
\begin{center} 
Arkady Vainshtein \\
\vspace{0.4cm}

{\normalsize
{\it  Theoretical Physics Institute, Univ. of Minnesota,
Minneapolis, MN 55455}}

\vspace{2cm}

{\large\bf Abstract}

\end{center}

\vspace*{.25cm}

The 't~Hooft model (two-dimensional QCD in the limit of large number
of colors) is used as a testground for calculations of nonleptonic and
semileptonic inclusive widths of heavy flavors based on the operator product
expansion (OPE). The OPE-based predictions up to terms ${\cal O}(1/m_Q^4)$,
inclusively, are confronted with the ``phenomenological" results, obtained by
summation of all open exclusive decay channels, one by one, a perfect
match is found. 

The issue of  duality violations is discussed, the amplitude of
oscillating terms is estimated. The method is applied to the realistic case of
hadronic $\tau$ decays.

\vspace{2cm}

\end{titlepage}

\section{Overview}
\label{sec:overview}

In this talk I will present recent results obtained in collaboration
with Ikaros Bigi, Mikhail Shifman and Nikolai Uraltsev~\cite{2d}. 
They refer to the applications of Wilson's operator product
expansion (OPE)~\cite{1} to inclusive decays of heavy
flavor hadrons.  The theory of such decays is at a rather advanced
stage now (see~\cite{optical} and references therein).

The decays of heavy flavor hadrons $H_Q$ are shaped by nonperturbative
dynamics.  While QCD at large distances is not yet solved,
considerable progress has been achieved in this problem. The inclusive 
widths are  expressed through  OPE. The
nonperturbative effects are then parameterized through expectation
values of various local operators ${\cal O}_i$ built from the quark
and/or gluon fields.  Observable quantities, such as total semileptonic and
nonleptonic widths of heavy hadrons $H_Q$, are then given by
\begin{equation}
\Gamma_{H_Q}= \frac{1}{M_{H_Q}}\sum_i {\rm Im}\, c_i(\mu)
\langle H_Q|{\cal
O}_i(\mu)| H_Q\rangle
\label{Gamma}
\end{equation}
where $c_i$ are the OPE coefficients, and $\mu$ stands for a
normalization point separating out soft contributions (which are
lumped into the matrix elements 
$\langle H_Q|{\cal O}_i(\mu)|H_Q\rangle$) 
from the hard ones (which belong to the coefficient
functions $c_i$).

There are a number of  issues, both conceptual and
technical, associated with the operator product expansion in QCD:
nonperturbative contributions to the OPE coefficients, dependence 
on normalization point, violations of local duality. 
These questions are circumvented in the so-called {\em practical
version} of OPE \cite{2} routinely used so far in all instances when
there is need in numerical predictions. This version is admittedly
approximate, however. The questions formulated above are legitimate, and
they deserve to attract theorists' attention, as they continue to cause
confusion in the literature.

We find it useful and instructive to study these subtle issues in
framework of the 't~Hooft model~\cite{H1} (see also Refs.~\cite{H2}).
The model is 1+1
dimensional QCD in the limit \mbox{$N_c\to\infty$}.  While
retaining basic features of QCD -- most notably quark confinement --
this mode is simpler without being trivial and can be solved
dynamically. 
The theory is superrenormalizable, i.e. very simple in the ultraviolet
domain.  As a result, the book-keeping of OPE becomes simple, and all
subtle aspects in the construction of the OPE can be studied in a
transparent environment.

Our study was motivated also by Ref.~\cite{5} where heavy flavor
inclusive widths were calculated numerically, by adding the exclusive
channels one by one.  It was found that the inclusive width
$\Gamma_{H_Q}$ approaches its asymptotic (partonic) value, and the sum
over the exclusive hadronic states converges rapidly.  At the same
time, small deviations from the asymptotic value observed in the
numerical analysis \cite{5} were claimed to be a signal of $1/m_Q$
corrections in the total width, in contradiction with the OPE-based
result.

We treated the very same problem {\em analytically}. The simplification
which allows the analytical solution comes from taking fermions $\psi$
(leptons or quarks) emitted in $Q \to q +\psi{\bar \psi}$ transition to
be massless. Calculating OPE coefficients and comparing the OPE
representation (\ref{Gamma}) with the sum over exclusive channels
which can be found from the 't~Hooft equation we observe a 
{\em perfect match} through order $(1/m_Q)^4$.

After testing the validity of OPE, we discuss the issue of violation
of local duality.  Oscillating contributions to the total width are
estimated.  They are suppressed by the high power of $1/m_Q$ which we
determined. Then we apply the same method in the real 1+3 QCD to
estimate duality violations in hadronic $\tau$ decays.

\section{Setup of the problem}
\label{sec:prelim}
\subsection{'t Hooft model}
\label{sub:model}
In two-dimensional QCD the Lagrangian looks superficially the same
as in four dimensions
\begin{equation}
{\cal L}_{1+1}=-\frac{1}{4g^2} \,
G_{\mu\nu}^a G_{\mu\nu}^a \,+\, \sum
\bar\psi_i
(i\not \!\!D -m_i)\psi_i \; , \; \;\;\;
i D_\mu=i\partial_\mu + A_\mu^a T^a\, ;
\label{11}
\end{equation}
$T^a$ denote generators of $SU(N_c)$ in the fundamental representation
and $\psi_i$ the quark field ($i$ is a flavor index) with a mass
$m_i$; $g$ the gauge coupling constant.

One has to keep the following peculiarities in mind: $g$ carries
dimension of mass as does $\bar \psi \psi$. With the theory being
superrenormalizable no (infinite) renormalization is needed;
observables like the total width $\Gamma_{H_Q}$ can be expressed in
terms of the {\em bare} masses $m_i$ and {\em bare} coupling $g$
appearing in the Lagrangian.

The 't~Hooft model is the $N_c \to \infty$ limit of 1+1 QCD.  A
parameter fixed in this limit is
\begin{equation}
\beta^2=\frac{g^2}{2\pi}\left( N_c -\frac{1}{N_c} \right)\;
\; \; {\rm with} \; \; \;
\lim_{N_c \to \infty} \beta ^2 = {\rm finite} \; .
\label{beta}
\end{equation}
This dimensionful quantity $\beta$ which provides an intrinsic mass
unit for the 't~Hooft model, can be seen as the analog of
$\Lambda_{\rm QCD}$ of four-dimensional QCD.

The heavy hadron $H_Q$ is the bound state of a heavy quark $Q$ and a
light spectator antiquark $q_{sp}$. The bound state is described by
the light-cone wave function $\phi(x),\;\; (x\in [0,1])$ which is the
eigenfunction of the 't~Hooft equation,
\begin{equation}
M^2_{H_Q}\varphi _{H_Q}(x) =
\left[
\frac{m_Q^2 - \beta ^2}{x} + \frac{m_{sp}^2 - \beta ^2}{1-x}
\right]
\varphi _{H_Q}(x) - \beta ^2 \int_0^1{\rm d}y \,\frac{\varphi
_{H_Q}(y)}{(y - x)^2}
\label{THOOFTIN}
\end{equation}
with boundary conditions $\varphi _{H_Q}(x=0)=\varphi _{H_Q}(x=1)=0$.

\subsection{Weak interaction}
\label{sub:weak}
Next we need to introduce a flavor-changing weak interaction;
we  choose it to be of the current-current  form:
\begin{equation}
{\cal L}_{\rm weak} =-\frac{G}{\sqrt{2}}\,(\bar q \gamma_\mu Q)
\,
(\bar{\psi}\gamma^\mu \psi )\;.
\label{lweak}
\end{equation}
The field $\psi$ can be
either the light quark or the lepton field to describe nonleptonic or
semileptonic decays, respectively.

For $N_c \to \infty$ factorization holds; i.e., the transition
amplitude can be written as the product of the matrix elements of the
currents $\bar q \gamma_\mu Q$ and $\bar{\psi}\gamma^\mu \psi$. 
For the inclusive widths which are discussed below the property of 
factorization can be expressed as follows:
\begin{eqnarray}
M_{H_Q}\Gamma_{H_Q}&=&{\rm Im} \int {\rm d}^2 x \, i \,
\langle H_Q|\,T\left\{ {\cal L}_{\rm weak}(x)
{\cal L}_{\rm weak}^\dagger(0)\right\}|H_Q\rangle \label{fact}\\
&=&G^2\, \int {\rm d}^2 x \, {\rm Im}\Pi_{\mu\nu}(x)\,  
{\rm Im}T^{\mu\nu}(x)\;,\nonumber\\
& & \nonumber\\
  \Pi_{\mu\nu}(x)&=& i\,\langle 0|T \left \{ \bar{\psi}(x) \gamma_\mu \psi(x)
\, \bar{\psi}(0) \gamma_\nu \psi(0)\right\}|0\rangle\,,\nonumber\\
T^{\mu\nu}(x)&=& i\,\langle H_Q |T \left \{ \bar{q}(x) \gamma^\mu Q(x)\,
\bar{Q}(0) \gamma_\nu q(0)\right\}|H_Q\rangle\,.\nonumber
\end{eqnarray}
This factorization follows from the fact that at $N_c\to \infty$ there is 
no communication between $ \bar{\psi}_a \gamma_\mu \psi_b$ and 
$ \bar{q} \gamma^\mu Q$ currents: any gluon exchange brings in 
a suppression factor $1/N_c^2$.

The only difference between the semileptonic and nonleptonic widths
resides in $\Pi_{\mu\nu}(x)$, in the first case $\psi$ is the leptonic field
while in the second case it is the quark field. 
At $m_\psi =0$, we get same $\Pi_{\mu\nu}(x)$ (up to the overall
normalization factor $N_c$). Indeed, for massless fermions
$\Pi_{\mu\nu}$ is given by  the well-known expression,
\begin{equation}
\Pi_{\mu\nu} (q)=\int {\rm d}^2 x \, e^{iqx} \Pi_{\mu\nu}(x)
=-\frac{1}{\pi }\, \left(\frac{q_\mu q_\nu }{q^2} - g_{\mu\nu}
\right)\, .
\label{piq}
\end{equation}
This expression obtained from a one-loop graph is known to be
exact. For this reason at $m_\psi =0$ the distinction between the
nonleptonic and semileptonic cases is actually immaterial.

A remarkable feature of Eq.~(\ref{piq}) is the occurrence of the pole
at $q^2=0$, which is specific for the vector interaction.  This means
that a pair of massless leptons or quarks produced by the vector
current is equivalent to one massless boson, whose coupling is
proportional to its momentum $q_\mu$. In the case of the quark fields,
it is known \cite{H2} from the early days of the 't~Hooft model that
the current $\bar{\psi} \gamma^\mu \psi$ produces from the vacuum only
one massless meson, the pion.

It means that we can use the substitution
$
  \bar{\psi}_a \gamma^\mu \psi_b \to \epsilon^{\mu\nu} \partial_\nu
\phi/\sqrt{\pi}
$
where the field $\phi$ describes a massless boson. For the $Q$ quark
decay width we have 
\begin{equation}
\Gamma_Q=\Gamma(Q \to \psi\;{\bar \psi}\;q)=\Gamma(Q \to \phi\;q)=
\frac{G^2}{4\pi}\cdot \frac{m_Q^2-m_q^2}{m_Q}\;.
\label{partonic}
\end{equation}

%%%%%%%%%%%
\section{Operator product expansion for inclusive widths}
\label{sec:OPE}
%%%%%%%%%%
The $1/m_Q$ expansion for inclusive widths of heavy flavor hadrons is
based on OPE for the weak transition operator
\cite{Vol}
\begin{equation}
 \int {\rm d}^2x\; i T\left\{ {\cal L}_{\rm weak}(x)
{\cal L}_{\rm weak}^\dagger (0)\right\} = \sum c_i(\mu){\cal
O}_i(\mu)
\, .
\label{5}
\end{equation}
The local operators ${\cal O}_i$ are ordered according to their
dimensions. The leading one is $\bar{Q}Q$ with dimension 
$d_{\bar{Q}Q}=1$.  
Higher operators have dimensions $d_i >1$. By dimensional
counting the corresponding coefficients are proportional to
$(1/m_Q)^{(d_i -2)}$. 
The coefficients $c_i$ (actually, we need Im$c_i$) are determined in perturbation
theory as a series in $\beta^2/m_Q^2$.  These coefficients are
saturated by the domain of virtual momenta $\sim m_Q$ and are infrared
stable by construction. All infrared contributions reside in the
matrix elements of the operators ${\cal O}_i$. 

At this point we see
 a drastic distinction between four- and two-dimensional
QCD.  In four dimensions the expansion parameter for the coefficients
is the running coupling $\alpha_s (m_Q )$; nonperturbative
contributions to the coefficients coming from distances $\sim 1/m_Q$
could, in principle, show up in the form $\exp (-C/ \alpha_s (m_Q ))
\sim (\Lambda_{\rm QCD}/m_Q )^\delta$ where $\delta$ is some positive
index.  In two-dimensional QCD such terms
cannot appear: an analog of the exponential term above would be
$\exp (- Cm_Q^2 /\beta^2 )$.

For the leading operator   $\bar{Q}Q$  its coefficient in the zero order in 
$\beta$ is ${\rm Im}c_{\bar{Q}Q}^0= \Gamma_Q/2$ where $\Gamma_Q$
is given by Eq.~(\ref{partonic}). We found  Im$c_{\bar{Q}Q}$ in all orders:
the only effect of high orders is the substitution $m_Q^2\to m_Q^2-\beta^2$ and
$m_q^2\to m_q^2-\beta^2$. This result is based on the nonrenormalization
theorem we proved for the $\bar{q} \gamma^\mu Q$ current at $q^2=0$.
For the next by dimension four-fermion operators ${\bar{Q}\Gamma
Q}{\bar{q}\Gamma q}$ we show that they will appear only in the $\beta^4$
order. By dimensional counting it means that their contribution to the total widths
is of order of $1/m_Q^5$.  

Putting everything together we  get the OPE representation
for the inclusive width:
\begin{equation}
\Gamma_{H_Q} = \Gamma_Q \left[
\frac{m_Q}{\sqrt{m_Q^2 - \beta ^2}}\, 
\frac{\langle H_Q|\bar QQ |H_Q \rangle}{2M_{H_Q}} +
{\cal O}\left(\frac{1}{m_Q^5}\right) \right] \, .
\label{GAMMADUAL}
\end{equation}
Note that dependence on $m_Q$ resides not only in the explicit OPE coefficient but
also in the matrix element. The relevant matrix element is 
\begin{equation}
\frac{\langle H_Q|{\bar Q} Q |H_Q
\rangle}{2M_{H_Q}}=1+ {\cal O}\left(\frac{1}{m_Q^2}\right)
\;,
\end{equation}
thus the absence of linear in $1/m_Q$
corrections is clear. We will prove a much stronger
statement in the next section: with
the accuracy of
$1/m_Q^5$ the OPE representation~(\ref{GAMMADUAL}) matches the
 summation of {\em exclusive} widths based on the 't~Hooft equation.

%%%%%%%%%%%%%
\section{Match between OPE-based expressions and  hadronic saturation}
\label{sec:match}
%%%%%%%%%%%%%
The exclusive width $\Gamma_n$ of two body decay $H_Q \to h_n +\phi$
where $h_n$ is the mesonic $[q {\bar q}_{sp}]$ state with the mass $M_n$ is
expressed via the overlap of the  initial and final 't~Hooft wave functions,
\begin{equation}
\label{excl1}
\Gamma _n =  \frac{G^2}{4\pi }
\frac{M_{H_Q}^2 - M_n^2}{M_{H_Q}}
\left|
\int _0^1 {\rm d}x \varphi _n(x)\varphi _{H_Q}(x)
\right|^2 \;.
\end{equation}
The wave function $\varphi _n(x)$ satisfies  the same 't~Hooft  equation
(\ref{THOOFTIN}) with evident substitutions, $Q \to q$, $M_{H_Q} \to M_n$.

Using this equation together with the completeness, 
$$
\sum \varphi_n(x)\varphi_n(y)=\delta (x-y)\;,
$$
we get a set of three sum rules for $\Gamma_n$:
\begin{eqnarray}
\frac{4\pi M_{H_Q}}{G^2}
\sum _{n=0}^{\infty}\frac{\Gamma _n}{M_{H_Q}^2 - M_n^2}
& = & 
\int _0^1 {\rm d} x \varphi _{H_Q}^2(x) = 1\;,
\label{SR1}\\
\frac{4\pi M_{H_Q}}{G^2}
\sum _{n=0}^{\infty}
\Gamma _n & = & 
(m_Q^2 - m_q^2) \int _0^1 \frac{{\rm d}x}{x} \,\varphi _{H_Q}^2(x)\;,
\label{SR2}\\
\frac{4\pi M_{H_Q}}{G^2}
\sum _{n=0}^{\infty}
\Gamma _n (M_{H_Q}^2 - M_n^2)
& = & 
(m_Q^2 - m_q^2)^2 \int _0^1 \frac{{\rm d}x}{x^2} \, \varphi _{H_Q}^2(x)\;.
\label{SR3}
\end{eqnarray}
Note that the sums runs over {\em all} states $h_n$, including those
unaccessible in the real decays of $H_Q$, i.e. with masses
$M_n>M_{H_Q}$. These transitions are still measurable by the process
of inelastic lepton scattering off the $H_Q$ meson. The first two sum rules 
were introduced in Ref.~\cite{Burkardt}.

It is not difficult to see that the next moment,
 $\sum _{n=0}^{\infty} \Gamma _n (M_{H_Q}^2 - M_n^2)^2$, is a
divergent sum because $\varphi _{H_Q}^2(x)/x^3$ would no longer be
integrable. It defines the asymptotics, $ \Gamma _n \propto 1/M_n^6 $, 
at large $n$. For this reason the ``unphysical'' part of the
sum~(\ref{SR2}) gives only $1/m_Q^5$ corrections and we derive from the
sum rule~(\ref{SR2})
\begin{equation}
\label{ghgq}
\Gamma_{H_Q}= \Gamma_Q \left[ \frac{m_Q}{M_{H_Q}}\int _0^1
\frac{{\rm d}x}{x}
\,\varphi _{H_Q}^2(x) + {\cal O}\left(\frac{1}{m_Q^5}\right) \right]\, .
\end{equation}
This expression coincides with the OPE result of Eq.~(\ref{GAMMADUAL})
as seen by rewriting the matrix
element in Eq.~(\ref{GAMMADUAL}) in terms of the ground state wave
function $\varphi_{H_Q}(x)$. (The factor $m_Q/(m_Q^2 -\beta^2)^{1/2}$
accounts for nonlogarithmic dependence of ${\bar Q} Q$ operator on
normalization point).

Thus, we have demonstrated the perfect match between OPE and the
hadronic saturation for the inclusive width.

%%%%%%%%%%%%
\section{Deviations from local duality}
\label{sec:violations}
%%%%%%%%%%%%%
Having established a perfect match between the OPE prediction for the
total width and the result of the saturation by exclusive decay modes,
through ${\cal O} (1/m_Q^4)$, we must now turn to the issue of where
the OPE-based prediction is supposed to fail.  The failure usually
goes under the name of ``duality violations", a topic under intense
scrutiny in the current literature.
 
We must precisely define what is meant by duality and its violations.
Sum rules relating certain moments of the imaginary part of transition
amplitude to matrix elements of consecutive terms in the OPE series
will be referred to as {\em global duality}.  Taken at their face
value, they are exact, to the extent of our knowledge of the coefficient
functions and matrix elements of the operators involved in OPE.
Therefore, it does not make any sense to speak about violations of the
global duality.

The notion of {\em local duality} on the other hand requires further
assumptions.  The difference between the OPE-based smooth
result and the experimental hadronic measurement is referred to as the
{\em duality violation} meaning the violation of {\em local duality}.
  Thus, the duality
violation is something we do not see in the (truncated) OPE series.
The duality-violating terms are exponential in the Euclidean domain
and oscillating (like $\sin[ (E/\Lambda_{\rm QCD})^k]$) in the Minkowski
domain.

\begin{figure}[h]
\epsfxsize=10cm
\centerline{\epsfbox{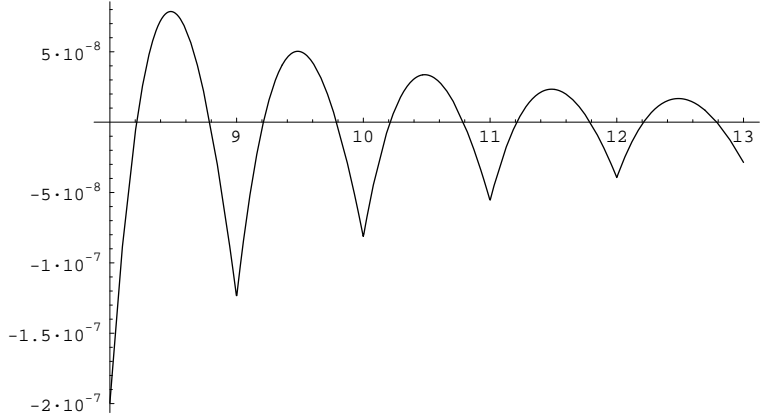}}
\caption{Oscillations in $\Gamma_{H_Q}$. The ratio $\Delta
\Gamma^{\mbox{\tiny osc}}/G^2 \beta$ is  presented as a
function of  $M_{H_Q}^2/\pi^2\beta^2$. }
\label{local}
\end{figure}

The appearance of duality violations in the form of oscillating terms
is evident in the 't~Hooft model where the spectral density is formed
by zero-width discrete states.  Indeed, each time a new decay channel
opens $d\Gamma_{H_Q} /d m_Q$ experiences a jump ($\Gamma_{H_Q}$
is continuous) , so that
immediately above the threshold, $d\Gamma_{H_Q} /d m_Q$ is larger than
the smooth OPE curve. In the middle between two successive thresholds
it crosses the smooth prediction, and immediately below the next
threshold $d\Gamma_{H_Q} /d m_Q$ is lower than the OPE-based
expectation. To estimate the amplitude of oscillations we use the
following model for exclusive widths $\Gamma_n$ at the range of $M_n$
in the vicinity of $M_{H_Q}$: 
\begin{equation}
\label{phi+n}
\Gamma_n = 3\pi G^2  \,\beta^7 \,\frac{M_{H_Q}^2 - M_n^2}{M_n^8}\;.
\end{equation}
Then for the oscillating part $\Delta \Gamma^{\mbox{\tiny osc}}$ of
the total width $\Gamma_{H_Q}$ as a function of $m_Q$ we get
\begin{equation}
  \label{oscillat}
\frac{\Delta \Gamma^{\mbox{\tiny osc}}}{\Gamma_Q}  = 6\pi^4
\left(\frac{\beta}{M_{H_Q}}\right)^9\,\left[x(1-x)-\frac{1}{6}\right]
\end{equation}
where
$$
x=\mbox{Fractional Part
of\/} \left(\frac{M_{H_Q}^2}{\pi^2\beta^2}\right)\,,\;\;\;\;x \in [0,1)\,.
$$
The plot of $\Delta \Gamma^{\mbox{\tiny osc}}$ is presented in 
Fig.~\ref{local}.
The amplitude of oscillations is $(3\pi^4/2)(\beta/M_{H_Q})^9$.
Thus, we see that violation of local duality is present but suppressed
as $1/m_Q^9$. If we could average over $m_Q$ in a sufficiently large
interval we would get an exponential suppression. Then it would be possible
to consider the OPE-based predictions beyond $1/m_Q^9$. 

%%%%%%%%%%%%%
\section{Duality violations in $\tau$ decays}
\label{sub:tau}
%%%%%%%%%%%%%

Let us discuss along similar lines a quantity of practical interest 
in 1+3 dimensions, namely the normalized hadronic
$\tau$ width $R_{\tau}$:
\begin{equation}
R_{\tau} \equiv
\frac{\Gamma ( \tau ^- \to \nu _{\tau}+{\rm hadrons})}
{\Gamma ( \tau ^- \to \nu _{\tau}e^- \bar \nu _e)}
= \int _0^{M_{\tau}^2} \frac{{\rm d}s}{M_{\tau}^2}
\left( 1-\frac{s}{M_{\tau}^2}\right) ^2
\left( 1+2\frac{s}{M_{\tau}^2}\right)
 \rho (s) 
\label{72}
\end{equation}
where 
$\rho(s)$ is the sum of spectral densities in the vector and axial-vector
channels.

To estimate the oscillating part of $R_{\tau}$ we consider the limit
of large $N_c$ and $M_\tau$. 
For large $N_c$ the spectrum of 1+3 QCD is expected to consist of an
infinite comb of narrow resonances -- in complete analogy to the
't~Hooft model \cite{largenc}.
To keep the closest parallel to it we further assume that
the high excitations are equally spaced in $m^2$.
This agrees with the general expectation of a string-like realization
of confinement leading to asymptotically linear Regge trajectories.
The masses of the excited states in, say, the vector channel are then
given by $m^2_n = m_{\rho}^2 + 2n/\alpha ^{\prime}$
\cite{Kprim}, with $\alpha ^{\prime}$ being the slope of the Regge
trajectory. Experimentally one finds
$2/\alpha ^{\prime} \simeq 2~$GeV$^2$.  

Then
spectral density $\rho (s)$ at large values of $s$ approaches the form:
\begin{equation}
\rho (s) =2 N_c \cdot \sum_{n=1}^\infty
\delta \left(\frac{s}{\sigma^2}
- n\right)\,;\qquad \sigma^2=\frac{2}{\alpha^{\prime}}\;,
\label{SPECTDENS}
\end{equation}
Equation~(\ref{SPECTDENS}) is clearly not expected to hold
at moderate and small values of $s$ where the
vector and axial-vector channels are drastically
different and the resonances are not equidistant.
However, details of the spectral densities at small $s$ play no role in
duality violation, they  change only the regular terms of the 
$1/M_\tau^2$ expansion.
The spectral density in Eq.~(\ref{SPECTDENS}) is dual to the
parton model result; i.e., it coincides with it after
averaging over energy,
$
\langle \rho (s)\rangle = 2N_c
$,
and $R_\tau$ approaches $N_c$ asymptotically.

\begin{figure}[h]
\epsfxsize=10cm
\centerline{\epsfbox{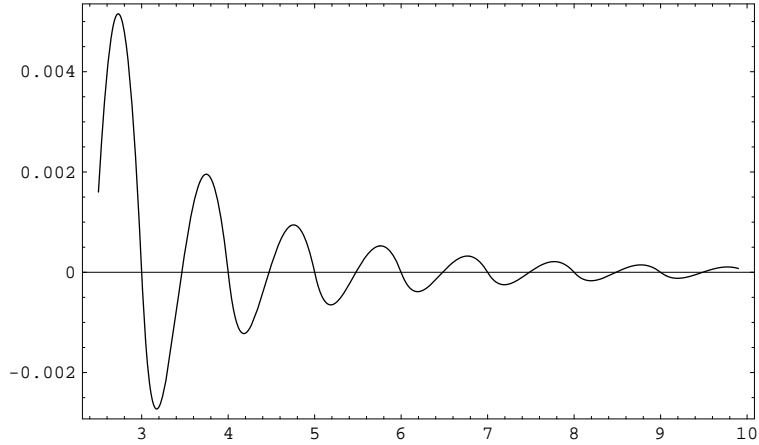}}
\caption{Oscillations in $R_\tau$. The plot of 
$\delta R^{\rm osc}/N_c$  is presented as 
a function of  $M_\tau^2/\sigma^2$.}
\label{oscillation}
\end{figure}

The sum over resonances in $R_\tau$ is  calculated analytically:
\begin{eqnarray}
R_{\tau} &\,=\,& R_{\tau}^{\mbox{\tiny OPE}} +
\delta R^{\rm osc}\;,\label{n9} \\
\frac{R_\tau^{\mbox{\tiny OPE}}}{N_c} &\,=\,& 1-\frac{\sigma^2}{M_\tau^2} +
\frac{1}{30}\left(\frac{\sigma^2}{M_\tau^2} \right)^4\;, \nonumber\\
\frac{\delta R^{\rm osc}}{N_c}&\;=\;& -\,
x(1-x)(1-2x)\, \left(\frac{\sigma^2}{M_\tau^2}\right)^3 \,+\,
\left[ x^2(1-x)^2-\frac{1}{30}\right]\,\left(\frac{\sigma^2}
{M_\tau^2}\right)^4,\nonumber
\end{eqnarray}
where 
$$
x={\rm Fractional~Part~of}\left(\frac{M_\tau^2}{\sigma^2}\right),
\;\;\;\: x\in [0,1)
\;.
$$
The result is a sum of two functions of $M_\tau^2$, the first 
one, $R_{\tau}^{\mbox{\tiny OPE}}$, is a smooth function expandable in
$1/M_\tau^2$, it coincides with the OPE 
prediction in the model.   The second one,
$\delta R^{\rm osc}$, oscillates with the period $\sigma^2$, 
see the plot of $\delta R^{\rm osc}/N_c$ in
Fig.~\ref{oscillation}.
The dominant component of $\delta R^{\rm osc}/N_c$  scales as 
$1/M_\tau^6$. 
This oscillating component vanishes at values of $M_\tau$ corresponding to the
new thresholds, and at one point in the middle between the successive
resonances; its second derivative  has a jump at the
thresholds, so we deal with a nonanalytical dependence on $M_\tau$.
The average of $\delta R^{\rm osc}$ vanishes while the amplitude of 
oscillations amounts to
\begin{equation}
\left| \frac{\delta R^{\rm osc}}{R_\tau}\right|_{\rm max}\;=\;
\frac{1}{3\sqrt{12}}\, \left(\frac{\sigma^2}{ M_\tau^2}\right)^3
\;.
\label{n12}
\end{equation}

Taking our estimate of the oscillation amplitude at its face value
and  using the actual value of the 
$\tau$ mass in Eq.~(\ref{n12}) we find $\delta R^{\rm osc}/R_\tau\sim 3\%$.
 One should not the number
take too literally  for many reasons: first of
all the $\tau$ mass is not much larger than the spacing between the
resonances, $\sigma^2/M^2_\tau \sim 2/3$; second, $N_c$ is not large 
enough to warrant the zero
width approximation, the nonvanishing widths lead to a further suppression of
deviations from duality.  Still we believe that the consideration is
instructive in a qualitative aspect.

\section*{Acknowledgments}
I am grateful to my collaborators I. Bigi, M. Shifman and N. Uraltsev.
This work was supported in part by the DOE under the grant number
DE-FG02-94ER40823.

\end{document}